\documentclass[a4,12pt, twocolumn,  usenatbib]{mn2e}
\usepackage[dvipdf]{graphicx, psfrag}
\usepackage{times}
\usepackage{color}
\usepackage{amsmath, amssymb}

\usepackage{natbib}
\usepackage{aas_macros}
\usepackage{ulem}

 \title[Magnetized stars with a differential field]
 {Magnetized stars with differential rotation  and  a differential toroidal field}
\author[K. Fujisawa]
{Kotaro Fujisawa
\thanks{E-mail: fujisawa@heap.phys.waseda.ac.jp}
\\
Advanced Research Institute for Science and Engineering, 
Waseda University, 3-4-1 Okubo, Shinjuku-ku, Tokyo 169-8555, Japan
}
\date{Accepted 2015 April 21. Received 2015 April 13; in original form 2015 February 25}

\def\Vec#1{\mbox{\boldmath $#1$}}
\def\D#1#2{\dfrac{d #1}{d #2}}

\def\P#1#2{\dfrac{\partial #1}{\partial #2}}

\begin{document}

\maketitle

\begin{abstract}
We have succeeded in obtaining magnetized equilibrium states
with differential rotation and {\it differential} toroidal magnetic fields. 
If an internal toroidal field of a protoneutron star 
is wound up from the initial poloidal magnetic field by differential rotation,
the distribution of the toroidal magnetic field is determined by the 
profile of this differential rotation. 
However, the distributions of the toroidal fields in 
all previous magnetized equilibrium studies 
do not represent the magnetic winding by the 
differential rotation of the star. In this paper, 
we investigate a formulation of a {\it differential} 
toroidal magnetic field that represents the 
magnetic field wound up by differential rotation.
We have developed two functional forms of {\it differential} 
toroidal fields which correspond to a 
$v$-constant and a $j$-constant field in analogy 
to differential rotations. 
As the degree of the {\it differential} becomes very high, 
the toroidal magnetic field becomes highly localized and 
concentrated near the rotational axis. 
Such a {\it differential} toroidal magnetic field would
suppress the low-$T/|W|$ instability 
more efficiently even if the total magnetic field 
energy is much smaller than that of a
non-{\it differential} toroidal magnetic field.
\end{abstract}

\section{Introduction}

A rapidly rotating protoneutron star are produced by a core collapse supernova,
accretion-induced collapse white dwarf. 
They can be subject to well-known non-axisymmetric instabilities (\citealt{Andersson_2003}).  
Stellar deformations caused by rotation and the non-axisymmetric
instabilities can generate strong gravitational wave signals, with
the global $m=2$ instabilities particularly related to 
gravitational wave production. Thus, a rapidly rotating protoneutron star
is a good source of gravitational radiation.

The development of non-axisymmetric instabilities is characterized 
by the energy ratio $\beta = T/ |W|$, where $T$ is the rotational energy 
and $W$ is the gravitational potential energy of the rotating star.
Studies of these instabilities were obtained 
in the nineteenth century (e.g., \citealt{Brayan_1889}).
Uniformly rotating incompressible fluids exhibit the $m=2$ f-mode
dynamical instability when the value of $\beta$ satisfies the threshold $\beta > 0.27$ 
(see also \citealt{Chandrasekhar_Ellipsoidal}).
On the other hand, there exists a secular
instability which is driven by dissipative 
processes, such as viscosity and gravitational radiation, 
which grows when the value of $\beta$ is larger than 0.14 
(see also numerical calculations by \citealt{Shibata_Karino_2004}). 
This threshold is easier to satisfy than the threshold for dynamical bar-mode 
instability, but the growth time-scale is much longer because the typical time-scale of the 
dissipation process is much larger than that of a dynamical time-scale (\citealt{Lai_Shapiro_1995}).
 
 The instability situation changes 
 if we introduce sufficiently large differential rotation.
 The linear analysis of a rapidly rotating star by \cite{Karino_Eriguchi_2003} 
 showed that the threshold of the dynamical instability decreases when the
 degree of the differential rotation is sufficiently large.
 Many numerical simulations have confirmed that $m=1$ or $m=2$ non-axisymmetric instabilities 
 can appear at a much lower $\beta$ than 0.27 if a sufficiently large differential 
 rotation is present (for example, $\beta \sim 0.14$ in \citealt{Centrella_et_al_2001}).
 This non-axisymmetric instability is referred to as 
 low-$T/|W|$ instability, and it can occur even 
 for $\beta \sim 0.01$ (\citealt{Shibata_Karino_Eriguchi_2002, Shibata_Karino_Eriguchi_2003};
 \citealt{Saijo_Baumugarte_Shapiro_2003}).
 Although numerical simulations have shown their existence, 
 the physical origin of low-$T/|W|$ instabilities
 remains unclear. The instabilities are associated with the presence of a 
 corotation resonance point inside the star (\citealt{Watts_Andersson_Jones_2005}; 
 \citealt{Saijo_Yoshida_2006}). The wave pattern speed of an unstable mode
 equals the local background fluid rotational
 velocity at the point. Corotation resonance has long been known to be a key factor 
 for instabilities in various astrophysical contexts, 
 such as the corotational instabilities for thin accretion discs, 
 the spiral pattern in galaxies (\citealt{Shu_1992}), 
 and the Papaloizou-Pringle instability for accretion tori 
 (\citealt{Papaloizou_Pringle_1984}; \citealt{Fu_Lai_2011a}). 
 By performing time evolutions of linear perturbations of 
 differential rotating stars,
 \cite{Passamonti_Andersson_2015} recently found 
 that the $\ell = m = 2$ f-mode becomes unstable as soon as a
 corotation point develops inside the star.
 The low-$T/|W|$ instabilities are thus a subclass of local shear instabilities. 

 The effect of magnetic fields is a significant issue concerning 
 these instabilities. Many numerical simulations have shown
 that low-$T/|W|$ instability may develop in some proto-neutron stars 
 with rapid rotation during 
 stellar core-collapses (\citealt{Ott_et_al_2005,Ott_et_al_2007}; \citealt{Shibata_Sekiguchi_2005a}) 
 because of their sufficiently large differential rotations.
 On the other hand, recent magnetohydrodynamical simulations have shown that the 
 dynamical bar-mode instability can be suppressed by magnetic 
 force (\citealt{Camarda_et_al_2009}; \citealt{Franci_et_al_2013}). 
 Since large toroidal fields can be wound up 
 from an initial poloidal magnetic field by differential rotation and
 amplified by magneto-rotational-instabilities 
 (\citealt{Akiyama_et_al_2003}; \citealt{Shibata_et_al_2006};
 \citealt{Obergaulinger_Aloy_Meuller_2006}; \citealt{Obergaulinger_et_al_2009};
 \citealt{Masada_et_al_2012};
 \citealt{Sawai_Yamada_Suzuki_2013}),
 rapidly rotating proto-neutron stars are expected to have large toroidal 
 magnetic fields, which would suppress the low-$T/|W|$ instability.

 Contrasting with the many numerical simulation studies, 
 \cite{Fu_Lai_2011} investigated the effect of a purely toroidal
magnetic field on the low-$T/|W|$ instability by applying linear analysis to a
cylindrical star with a simple model of the toroidal magnetic field 
(\citealt{Ostriker_1965}; \citealt{Saijo_Yoshida_2006}).
They found that the toroidal magnetic field suppresses the 
low-$T/|W|$ instability when the total toroidal magnetic energy
${\cal M}$ is in the order of 0.2 $T$ or larger. Such magnetic energy 
corresponds to toroidal fields of a few $10^{16}$ G in their magnetic field 
model. This study also found that the corotation point 
is split into two slow magnetosonic resonances 
in the presence of strong toroidal fields. 
Therefore, the unstable mode in the presence of the toroidal field is 
determined by both corotation and slow magnetosonic resonances 
inside the magnetized star.  

Very recently, however, \cite{Muhlberger_et_al_2014} 
performed accurate general relativistic magnetohydrodynamical 
simulations and found that the low-$T/|W|$ 
instability would be suppressed by a weaker magnetic field energy than given in \cite{Fu_Lai_2011}. 
In their simulation models, the large toroidal magnetic field ($\sim 10^{16} $G) 
is wound up from the initial poloidal field ($\sim 10^{14}$G) by differential rotation.
Since the toroidal magnetic fields are wound up in their models, 
the distributions of the toroidal magnetic field are determined by the profile of the 
differential rotation. The distributions of the toroidal magnetic fields are different from
the simple model used in \cite{Fu_Lai_2011}.
The maximum strength of the local toroidal magnetic field reached approximately $10^{17}$ G in
 their simulations, but the total magnetic energy peaked at $0.56 \%$ of the
 kinetic energy $T$. Although this energy ratio (${\cal M}/T \sim 5.6 \times 10^{-3}$) 
 is much smaller than that in \cite{Fu_Lai_2011} (${\cal M}/T \sim 0.2$),
 the low-$T/|W|$ instability was stabilized by the toroidal magnetic fields 
 in their simulation models. They also showed that slow magnetosonic resonances 
 appear in their simulation models (fig.  10 in their paper).
 This result shows that both the total magnetic energy and 
 the structure of the toroidal field are important for suppression by magnetic fields
 because low-$T/|W|$ instabilities significantly depend 
 on the values of the local rotational velocity (degree of the differential rotation;
 \citealt{Shibata_Karino_Eriguchi_2003}).
 \cite{Fu_Lai_2011}, however, calculated only one simple toroidal 
 magnetic field model which did not represent the wound magnetic field.
 Therefore, we must consider various toroidal magnetic fields in order 
 to study the linear analysis and address the suppression systematically. 
 
 Although many equilibrium states with purely toroidal magnetic fields have 
been consistently obtained in both Newtonian 
 (e.g., \citealt{Miketinac_1973}; \citealt{Lander_Jones_2009}; \citealt{Fu_Lai_2011}) and general 
 relativistic (e.g., \citealt{Kiuchi_Yoshida_2008}; \citealt{Yasutake_Kiuchi_Kotake_2010};
 \citealt{Frieben_Rezzolla_2012}; \citealt{Pili_et_al_2014}) frameworks, 
 all of  their toroidal magnetic field configurations are essentially 
 same-class solutions. When we calculate a stationary axisymmetric and barotropic rotating star
 with purely toroidal magnetic fields, we obtain two arbitrary functions 
 from the integrability condition of  the magnetohydrodynamic equation that
 are related to the rotation law of the 
 angular velocity and the distribution of the toroidal magnetic field. 
 Rotating equilibrium states have been studied and many rotation laws 
 have been investigated in both Newtonian and general relativistic frameworks 
 (e.g., \citealt{Eriguchi_Muller_1985}; \citealt{Hachisu_1986a};
 \citealt{Komatsu_Eriguchi_Hachisu_1989a}; \citealt{Galeazzo_Yoshida_Eriguchi_2012}).
 In contrast, only a few works have studied purely toroidal 
 magnetized equilibrium states, Almost all previous works have employed 
 only a simple functional form
 (\citealt{Miketinac_1973};  \citealt{Kiuchi_Yoshida_2008}; 
 \citealt{Lander_Jones_2009}; \citealt{Yasutake_Kiuchi_Kotake_2010};
\citealt{Fu_Lai_2011}; \citealt{Frieben_Rezzolla_2012}; \citealt{Pili_et_al_2014}).
 The functional form of a toroidal magnetic field is considered {\it rigid} in analogy to
 the rotation law. Although a {\it rigid} toroidal field is simple and widely accepted,
 the distributions of the toroidal field do not represent those wound up by differential
 rotation. Thus, we need to construct a {\it differential } toroidal magnetic field 
 which represents magnetic winding by differential rotation and
 to treat the linear analysis and the suppression more systematically. 
 As a first step, this paper  investigate new functional forms of a {\it differential} field 
 and obtains new solutions of purely toroidal magnetized equilibria. 
 These solutions will be useful for equilibrium models for linear analysis.

 The remainder of this paper is organized as follows. In Section 2, we describe 
 the new functional forms of the arbitrary functions. In Section 3, 
 we present results from numerical calculations. Discussion
 and conclusions are given in Section 4.


\section{Basic equations and formulations}

We employed the formulation developed by \cite{Kiuchi_Yoshida_2008} and 
\cite{Lander_Jones_2009}. We will first briefly summarize the formulation below
while the remainder of the section focuses on explaining the new functional forms in detail.
Both cylindrical coordinates ($\varpi$, $\varphi$, $z$) and 
spherical coordinates ($r$, $\theta$, $\varphi$) are utilized in this paper.

\subsection{Assumptions and basic equations}

As per previous works, 
we make the following assumptions for magnetized stars.
\begin{enumerate}
 \item The system is in a stationary state, i.e. $\partial/\partial t = 0$.
 \item The rotational axis and magnetic axis coincide. 
 \item The configuration is axisymmetric about the rotational axis,
       i.e., $\partial / \partial\varphi = 0$.
 \item There are no meridional flows or poloidal magnetic fields, i.e.
       the magnetic field is purely toroidal. 
 \item The star is self-gravitating.
 \item The system is treated in the framework of non-relativistic physics.
 \item No electric current is assumed in the vacuum region.
 \item There is no current sheet, i.e. the magnetic field
       continues smoothly.
 \item The barotropic equation of state is assumed.
\end{enumerate}
These assumptions are adopted here for simplicity.
For a barotropic equation of state, we use a 
polytropic equation of state
\begin{eqnarray}
 P = K \rho^{1+1/N},
\end{eqnarray}
where $P$ and $\rho$ are the gas pressure and the density, 
respectively, and $K$ and $N$ are constants.  

Under these assumptions, 
the equations that describe this system are the Euler equations
\begin{eqnarray}
 \frac{1}{\rho} \nabla p = - \nabla \phi_g + \Vec{F} + \Vec{L},
\label{Eq:eular}
\end{eqnarray}
\begin{eqnarray}
 \Vec{F} = \frac{v_\varphi^2}{\varpi} \, \Vec{e}_\varpi + 0 \, \Vec{e}_\varphi + 0 \, \Vec{e}_z,
\end{eqnarray}
\begin{eqnarray}
 \Vec{L} = \frac{1}{\rho} \left( \frac{\Vec{j}}{c} \times \Vec{B}  \right) =
 \frac{1}{\rho}\left\{ 
- \frac{j_z}{c}B_\varphi  \Vec{e}_\varpi + 0 \, \Vec{e}_\varphi + \frac{j_\varpi}{c} B_\varphi \Vec{e}_z
\right\},
\end{eqnarray}
the integral form of the Poisson equation
\begin{eqnarray}
\phi_g(\Vec{r}) = - G \int \frac{\rho(\Vec{r}')}{|\Vec{r} - \Vec{r}'|} \, dV'
\end{eqnarray}
and stationary Maxwell's equations,
\begin{eqnarray}
 \nabla \cdot \Vec{B} = 0,
\end{eqnarray}
\begin{eqnarray}
 \nabla \times \Vec{B} = 4 \pi \frac{\Vec{j}}{c},
\end{eqnarray}
\begin{eqnarray}
 \nabla \times \Vec{E} = 0,
\end{eqnarray}
 where $\phi_g$, $v_\varphi$, $\Vec{j}$, $c$, $B_\varphi$, $G$ and $\Vec{E}$
are, respectively, the gravitational potential, the toroidal velocity, the current density,
the speed of light, the toroidal magnetic field, the gravitational constant, and 
the electric field.  Using these equations,  we obtain the first integral of 
the Euler equation (see e.g., \citealt{Lander_Jones_2009}) as
\begin{eqnarray}
 \int \frac{dp}{\rho} = - \phi_g + \phi_R - \phi_M + C,
\end{eqnarray}
\begin{eqnarray}
\phi_R = \int \Omega^2(\varpi) \varpi \, d\varpi,
\end{eqnarray}
\begin{eqnarray}
 \phi_M = \frac{1}{4\pi}\int \frac{I(\lambda)}{\lambda} \D{I(\lambda)}{\lambda} \, d\lambda,
\end{eqnarray}
where $C$ is an integral constant, and  $\phi_R$ and $\phi_M$ are 
the rotational potential and  the magnetic potential respectively.
$\Omega(\varpi)$ and $I(\lambda)$ are arbitrary functions of $\varpi$ and $\lambda$, 
respectively, and $\lambda$ is a scalar 
function defined by $\lambda \equiv \rho \varpi^2$. 
These arbitrary functions are related to the toroidal velocity and 
the toroidal magnetic field according to
\begin{eqnarray}
 v_\varphi = \varpi \Omega(\varpi) ,
\end{eqnarray}
\begin{eqnarray}
 B_\varphi = \frac{I(\lambda)}{\varpi}.
\end{eqnarray}
Therefore, there appears to be two arbitrary functions 
when we calculate a rotating star with a purely toroidal magnetic field.
Note that we can choose any kind of functional forms of these potentials 
respectively under the assumption of no poloidal magnetic field and no meridional flow, 
because the toroidal magnetic field does not produce a force 
that acts directly onto the rotation and the rotation does not change the 
toroidal magnetic field.
We can obtain one rotating magnetized equilibrium solution 
by fixing these two functional forms.
The arbitrary function $\Omega$ corresponds 
to the rotation law of the rotating star. We can obtain various differential 
rotating stars by changing the functional form of $\Omega$.
On the other hand, the physical meaning of the arbitrary function $I$ is less clear.
In all previous works, a simple functional form was adopted with no
investigation of  new functional forms that express
various toroidal magnetic fields such as differential rotations. 
This paper discusses new functional forms of $I$ that correspond to
the toroidal magnetic field wound up by the differential rotation.

\subsection{Functional forms of arbitrary functions and their physical meanings}

First, we will review functional forms of $\Omega$ and 
discuss the rotation laws. The following three types of rotation law are usually 
used (see e.g., \citealt{Eriguchi_Muller_1985}; \citealt{Hachisu_1986a}):
\begin{enumerate}
 \item rigid rotation:
\begin{eqnarray}
\Omega^2(\varpi) =  \Omega_0^2,
\end{eqnarray}
 \item $v$-constant rotation:
\begin{eqnarray}
 \Omega^2(\varpi) = \frac{v_0^2}{(\varpi^2 + d^2)},
\end{eqnarray}
 \item $j$-constant rotation:
\begin{eqnarray}
 \Omega^2(\varpi) = \frac{j_0^2}{ (\varpi^2 + d^2)^2},
\end{eqnarray}
\end{enumerate}
where $\Omega_0$, $v_0$, $j_0$, and $d$ are constants.
$d$ represents the degree of the differential rotation.  
As $d$ increases for the latter two laws, their 
differential rotations approach rigid rotation.
As $d$ decreases towards zero, the latter two rotation laws approach
the true rotation denoted by their respective names. That is, 
$v_\varphi$ becomes strictly constant in space for the second law
($v$-constant rotation) and the specific 
angular momentum per unit mass, 
$j = \varpi^2 \Omega$, becomes exactly constant in space for the third law
($j$-constant rotation).
For these cases, we can obtain the explicit forms of the rotational potentials 
as follows.
\begin{enumerate}
 \item For rigid rotation:
       \begin{eqnarray}
	\phi_R(\varpi) =  \frac{\Omega_0^2 \varpi^2}{2},
       \end{eqnarray}
 \item For $v$-constant rotation:
       \begin{eqnarray}
	\phi_R(\varpi) = \frac{v_0^2 \ln (\varpi^2 + d^2)}{2}, 
       \end{eqnarray}
 \item For $j$-constant rotation:
       \begin{eqnarray}
	\phi_R (\varpi) = -\frac{j_0^2}{ 2 (\varpi^2 + d^2)} .  
       \end{eqnarray}
\end{enumerate}
These types of rotation laws have also been studied 
in general relativistic frameworks
(e.g., \citealt{Komatsu_Eriguchi_Hachisu_1989a, Komatsu_Eriguchi_Hachisu_1989b}).
Recently, \cite{Galeazzo_Yoshida_Eriguchi_2012} studied 
rotation laws systematically and obtained a new functional form,
\begin{eqnarray}
 \Omega(\varpi) = \Omega_c \left\{ 1 + \left( \frac{\varpi}{\varpi_0} \right)^2  \right\}^{{1}/{\gamma}},
\end{eqnarray}
where $\Omega_c$ and $\gamma$ are constants. We can invoke a wide range of 
rotation laws by changing the value of $\gamma$ (\citealt{Galeazzo_Yoshida_Eriguchi_2012}).
Almost all works concerning the low-$T/|W|$ instability 
have applied the  $j$-constant functional form for differential rotations of initial models 
(e.g., \citealt{Shibata_Karino_Eriguchi_2002,Shibata_Karino_Eriguchi_2003}).

In contrast, only the following simple power-law type functional form 
of $I$ has been adopted in  previous works 
(\citealt{Kiuchi_Yoshida_2008}; \citealt{Lander_Jones_2009}; \citealt{Yasutake_Kiuchi_Kotake_2010};
\citealt{Frieben_Rezzolla_2012}; \citealt{Pili_et_al_2014})
\begin{eqnarray}
 I(\lambda) = I_0 \lambda^k,
\label{Eq:Bt_rigid}
\end{eqnarray}
where $k$ is a constant. Since we have assumed that the magnetic field continues smoothly,
the regularity of the toroidal magnetic field is required.
The parameter $k$ must be $k \geq 1$, because
the functional form with $k \geq 1$ satisfies the regularity condition 
at the centre ($B_\varphi \rightarrow 0$, $\varpi \rightarrow 0$). 
We can derive the magnetic potential using this power-law functional form as 
\begin{eqnarray}
 \phi_M = \frac{I_0^2}{4\pi} k \int \lambda^{2k-2} d\lambda &=& \frac{I_0^2}{4\pi} \frac{k}{2k-1} \lambda^{2k-1} \nonumber \\
  &=& \frac{I_0^2}{4 \pi} \frac{k}{2k-1} (\rho \varpi^2)^{2k-1}.
\end{eqnarray}
As we have seen, the exponents of the rotational potential are
$\phi_R \sim \varpi^2$ (rigid), $\phi_R \sim \varpi^{-1}$ ($v$-constant), 
and $\Phi_R \sim \varpi^{-2}$ ($j$-constant).
The exponents of rotational potentials of the differential rotations are
 negative, while that of rigid rotation is positive.
The exponent of the rotational potential characterize the rotation law.
On the other hand, the magnetic potential becomes 
$\phi_M \sim \varpi^{4k - 2} (k \geq 1)$ when the star is incompressible (\citealt{Kiuchi_Yoshida_2008}).
Since $k$ is limited to $k \geq 1$ by the boundary condition, 
the exponent of this magnetic potential is always positive. 
Therefore, we consider this functional form of magnetic potential as a {\it rigid} 
toroidal magnetic field in analogy to the rotation law.

If the toroidal magnetic field is wound up from the initial 
poloidal fields by the differential rotations, 
the distribution of the toroidal magnetic field 
is determined by the  profile of this differential rotation.
If we neglect the meridional flow,  
the evolution of the toroidal magnetic field is described by the 
following induction equation:
\begin{eqnarray}
 \P{B_\varphi}{t} = \left\{  \nabla \times\left( \Vec{v} \times \Vec{B} \right) \right\}_\varphi 
  \sim  \P{}{z} (v_\varphi B_z) + \P{}{\varpi} (v_\varphi B_\varpi).
\end{eqnarray}
If the initial magnetic field is purely poloidal ($B_p$) and 
$B_p$ only has a constant $\varpi$-component 
($\Vec{B}_p = B_0 \, \Vec{e}_\varpi + 0 \,   \Vec{e}_\varphi + 0 \,  \Vec{e}_z$),
the poloidal field is wound up by the differential rotation and the 
profiles of the toroidal magnetic field become
\begin{eqnarray}
 B_\varphi \sim B_0 \P{}{\varpi} (\varpi \Omega) \sim
\left\{
\begin{array}{cc}
 \varpi^{-1} & (v \rm{\, -const.}), \\
 \varpi^{-2} & (j \mathrm{\, -const.}).
\end{array}
\right.
\label{Eq:Bt_wound}
\end{eqnarray} 
Therefore, the toroidal magnetic field wound up by the differential rotation 
must satisfy the magnetic winding in equation (\ref{Eq:Bt_wound}), 
but we have not obtained such toroidal magnetic field distributions 
by using a {\it rigid} toroidal field (equation \ref{Eq:Bt_rigid}).
Since the rotation and the toroidal magnetic field are not
necessarily consistent with each other under the assumption 
of no meridional flow and no poloidal magnetic field, 
we can select the functional forms of $I(\lambda)$ that are 
most compatible with the chosen rotational potential.

Here, we  introduce two new functional forms of $I$ whose toroidal 
magnetic fields satisfy equation (\ref{Eq:Bt_wound}).
We call the functional forms {\it differential} toroidal fields in
analogy to rotation law and rotational potentials. 
The functional form of a type 1 {\it differential } toroidal field is
\begin{eqnarray}
 I(\lambda) = I_0 \frac{\lambda}{\sqrt{\lambda + h^2}}
\end{eqnarray}
and the functional form of a type 2 {\it differential} toroidal field is
\begin{eqnarray}
 I(\lambda) = I_0 \frac{\lambda}{\lambda + h^2},
\end{eqnarray}
where $h$ is a constant. $h$ represents the degree of the {\it differential}
toroidal magnetic field. The magnetic potentials of these functions are
\begin{eqnarray}
 \phi_M(\lambda) = \frac{I_0^2}{4\pi} \left\{ \frac{\ln ( \lambda + h^2 )}{2}  -  \frac{h^2}{2 (\lambda + h^2)}\right\},
\end{eqnarray}
for a type 1 {\it differential} magnetic potential and
\begin{eqnarray}
 \phi_M(\lambda) = - \frac{I_0^2}{4\pi} \frac{h^2}{2 \left(\lambda + h^2 \right)^2},
\end{eqnarray}
for a type 2 {\it differential} magnetic potential. 
Since the exponents of these {\it differential} magnetic potentials 
are $\sim \varpi^{-2}$ in type 1 and $\sim \varpi^{-4}$ in type 2, 
we can obtain a wide range of the distributions of the 
toroidal magnetic field using these functional forms,
including distributions which meet
the condition of equation (\ref{Eq:Bt_wound}) 
(see the profiles of toroidal magnetic fields in fig. \ref{Fig:profiles}).

\subsection{Numerical settings and accuracy check}

Following  previous works (\citealt{Kiuchi_Yoshida_2008}; \citealt{Lander_Jones_2009}),
we adopted Hachisu's self-consistent scheme 
(\citealt{Hachisu_1986a}; \citealt{Hachisu_1986b}),  which is a very 
powerful numerical scheme for obtaining a rotating star in equilibrium. 
In this scheme, we fix the axis ratio $q \equiv r_{pol.} / r_e$ 
to obtain the equilibrium configuration, where $r_{\rm pol.}$ and $r_{\rm e}$ 
are the polar radius and the equatorial radius, respectively.  
When we calculate magnetized rotating equilibria, 
we need to fix another parameter. We choose $\hat{I}_0$ in the arbitrary functions 
as the parameter in this paper. Thus, we fixed both $q$ and $\hat{I}_0$
during our numerical iteration cycles and obtained one equilibrium state after the iteration.

The physical quantities are transformed to be dimensionless in actual numerical computations.
We adopted the dimensionless forms in \cite{Fujisawa_Eriguchi_2013} by
utlising the maximum density $\rho_{\max}$, the maximum pressure $p_{\max}$,
and the equatorial radius $r_e$ as follows:
\begin{eqnarray}
 \hat{r} \equiv \frac{r}{r_e} = \dfrac{r}{\sqrt{\dfrac{1}{\alpha} \dfrac{p_{\max}}{4\pi G \rho^2_{\max}}}},
\end{eqnarray}
\begin{eqnarray}
 \hat{\rho} \equiv \frac{\rho}{\rho_{\max}},
\end{eqnarray}
where $\alpha$ is introduced to ensure a unity distance from 
the centre to the equatorial surface of the star. 
Other physical quantities are also transformed to be dimensionless 
(see the Appendix in \citealt{Fujisawa_Eriguchi_2013}). 
A term with a hat ($\hat{}$) denotes a dimensionless quantity. 

To see the global characteristics of magnetized rotating equilibria, 
we define some integrated quantities as follows:
\begin{eqnarray}
 W = \frac{1}{2}\int \rho \phi_g \, dV,
\end{eqnarray}
\begin{eqnarray}
 T = \frac{1}{2} \int \rho (\varpi \Omega)^2 \, dV,
\end{eqnarray}
\begin{eqnarray}
 \Pi = \int p \, dV,
\end{eqnarray}
\begin{eqnarray}
 U  = N \Pi,
\end{eqnarray}
\begin{eqnarray}
 {\cal M} = \frac{1}{8\pi} \int B_\varphi^2 \, dV,
\end{eqnarray}
where $W$, $T$, $\Pi$, $U$, and ${\cal M}$ are the gravitational energy, the rotational energy, 
the total pressure, the internal energy, and the magnetic energy respectively.
We also define the volume averaged magnetic field  $B_{\rm ave}$ (\citealt{Fujisawa_Yoshida_Eriguchi_2012}) as
\begin{eqnarray}
 B_{\rm ave} \equiv \sqrt{\frac{\int B_\varphi^2 \, dV}{\int \, dV}}
= \sqrt{\frac{8\pi {\cal M}}{V}},
\end{eqnarray}
where $V$ is the volume of the star. This term is useful when we evaluate 
the localization of the magnetic field. 
We also calculate the toroidal Alfv\'en velocity as
\begin{eqnarray}
 V_A = \frac{B_\varphi}{\sqrt{4 \pi \rho}}.
\end{eqnarray}
Note that we checked the numerical convergence by using the Virial theorem (Appendix A)
and ensured a sufficient number of mesh points for numerical computations in this paper.

\section{Numerical results}

This section shows our numerical solutions with
differential rotation and a {\it differential} toroidal field.
We fixed the polytropic index $N=1$ as a simple equation 
of state for a proto-neutron star. 
We adopted the $j$-constant rotation law with $\hat{d} = 0.3$ 
as the differential rotation. 
The low-$T/|W|$ instability of this differential rotating star 
commences  when the value of $\beta$ becomes larger than $0.03$ (\citealt{Shibata_Karino_Eriguchi_2003}).
We fixed the axis ratio $q = 0.8$ and obtained the rotating equilibrium state
whose $\beta$ is 0.04. Panel (a) in fig. \ref{Fig:contours} is a density contour
of this rotating star without toroidal magnetic fields. 
The isopycnic surfaces are distorted by the strong rotation.
This rotating star gains the low-$T/|W|$ instability
if it does not have a toroidal magnetic field. 

To show the profiles of the re-dimensionalized toroidal magnetic fields, 
we used a neutron star's typical values of mass ($M = 1.4 M_\odot$) 
and central density ($\rho_{\max} = 10^{15} {\rm g/cm^3}$).
The maximum strength of the magnetic field is approximately a few times 
$10^{16}$ G and the maximum value of the Alfv\'en velocity is 
a few times $10^8 \mathrm{cm/s}$ in this model.

\subsection{{\it Differential} toroidal magnetic field}

\begin{table*}
\begin{center}
\caption{Numerical results for each model. All solutions have almost the same 
magnetic energy. }

 \begin{tabular}{cccccccccc}
\hline  
 $q$ & $\hat{I}_0^2$ &  $k$ or $\hat{h}$  & $\beta$ & ${\cal M} / T$ 
& $B_{\max} (G)$ & $B_{\max} / B_{\rm ave}$ & $V_{A\max}$ $(\mathrm{cm/s})$  & VC \\
\hline
&&&&& \hspace{-100pt} No magnetic \\
\hline
0.8 & 0      & - & 4.00E-2 & 0.00E0 & 0 & -  & - & 1.11E-5 \\
\hline
&&&&& \hspace{-100pt} {\it Rigid } toroidal \\
\hline
 0.8 & 7.12E-3 & $k=1$ & 4.02E-2 & 1.00E-2 & 3.67E16 & 1.88 & 4.14E8 & 1.12E-5 \\
 0.8 & 2.95E-1 & $k=2$ &  4.02E-2  & 1.00E-2 & 4.84E16 & 2.48 & 5.78E8 & 1.12E-5  \\
\hline
 &&&&& \hspace{-100pt} Type 1 {\it differential } toroidal \\
\hline
0.8 & 9.27E-4 & $\hat{h} =0.1$ & 4.02E-2 & 1.00E-2 & 3.30E16 & 1.70 & 3.15E8 & 1.12E-5 \\
\hline
&&&&& \hspace{-100pt} Type 2 {\it differential } toroidal \\
\hline
0.8 & 3.32E-4 & $\hat{h}=0.3$ & 4.02E-2 & 1.00E-2 & 3.50E16 & 1.80 & 3.20E8 & 1.12E-5\\
0.8 & 1.72E-4 & $\hat{h} =0.2$ & 4.02E-2 & 1.00E-2 & 3.85E16 & 1.99 & 3.45E8 & 1.12E-5\\
0.8 & 8.30E-5 & $\hat{h}=0.1$ & 4.03E-2 & 1.00E-2 & 5.39E16 & 2.77 & 4.80E8 & 1.15E-5\\
\hline
 \end{tabular}
\label{Tab:tab1}
\end{center}
\end{table*}

\begin{figure*}
 \includegraphics[width=8.2cm]{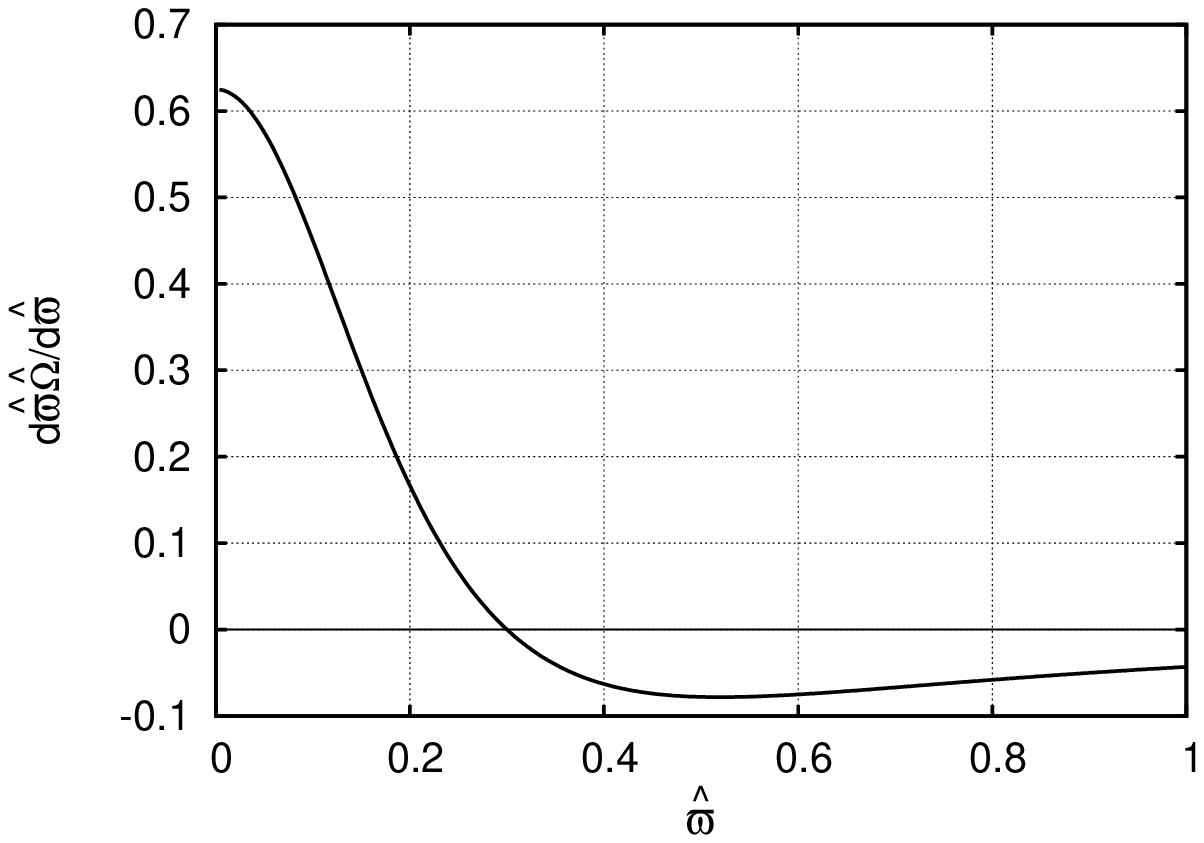}
 \includegraphics[width=8.8cm]{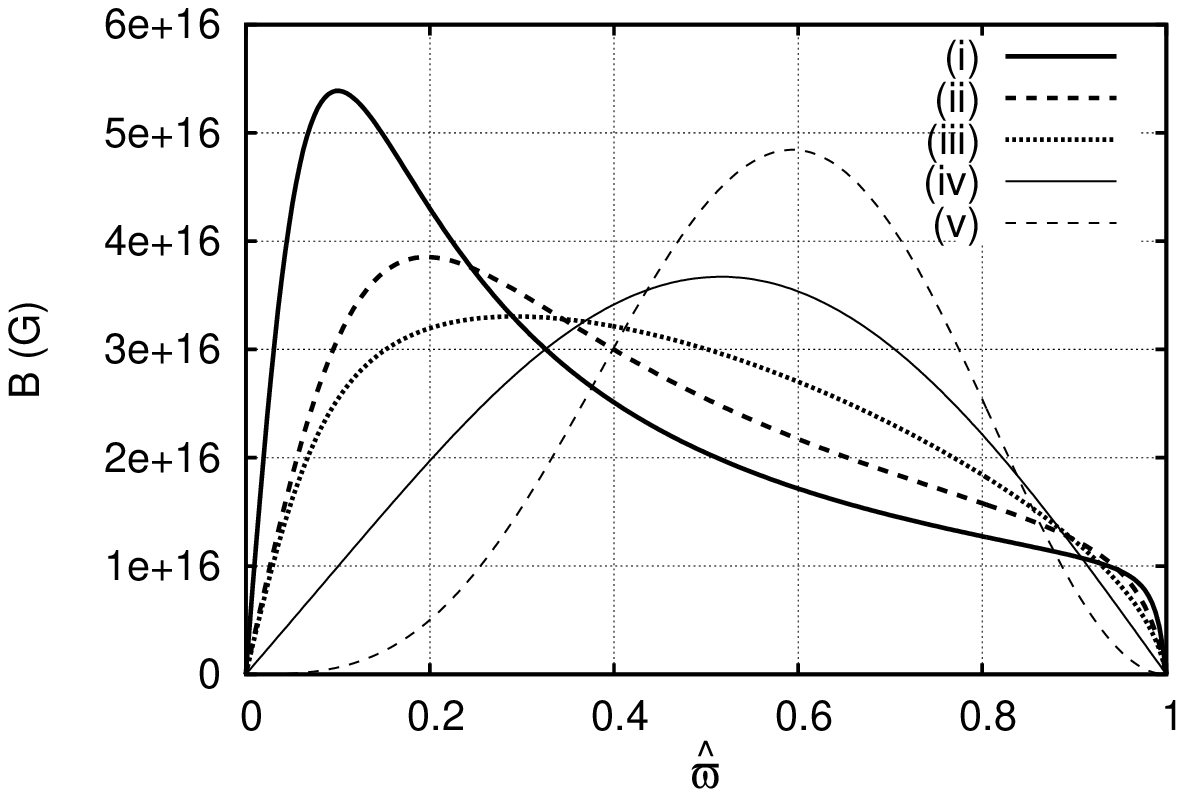}

\caption{Left: the profile of $d\hat{\varpi}\hat{\Omega} / d\hat{\varpi}$
with $\hat{d} = 0.3$ $j$-constant rotation.
Right: the profiles of $B_\varphi$ on the equatorial plain.
The vertical axis denotes the re-dimensionalized toroidal magnetic field in units of Gauss.
(i) Type 2 {differential} toroidal field with $\hat{h} = 0.1$. 
(ii) Type 2 {differential} toroidal field with $\hat{h} = 0.2$.
(iii) Type 1 {differential} toroidal field with $\hat{h} = 0.1$.
(iv) {Rigid} toroidal field with $k = 1$.
(v) {Rigid} toroidal field with $k = 2$.
} 
\label{Fig:profiles}
\end{figure*}
 
\begin{figure*}
\begin{tabular}{cc}
 \includegraphics[width=8.5cm]{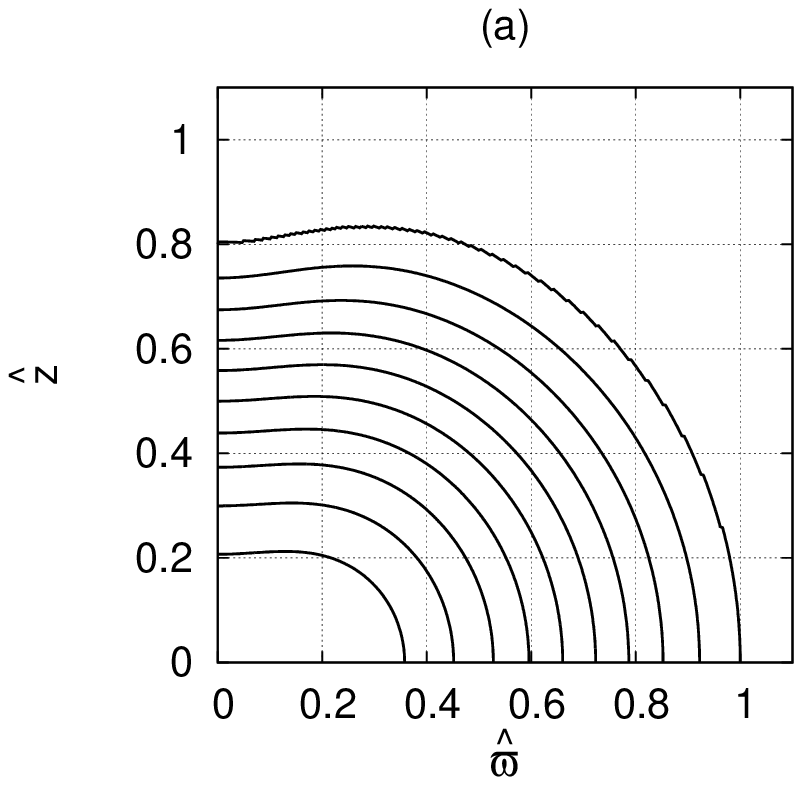}  &
\includegraphics[width=8.5cm]{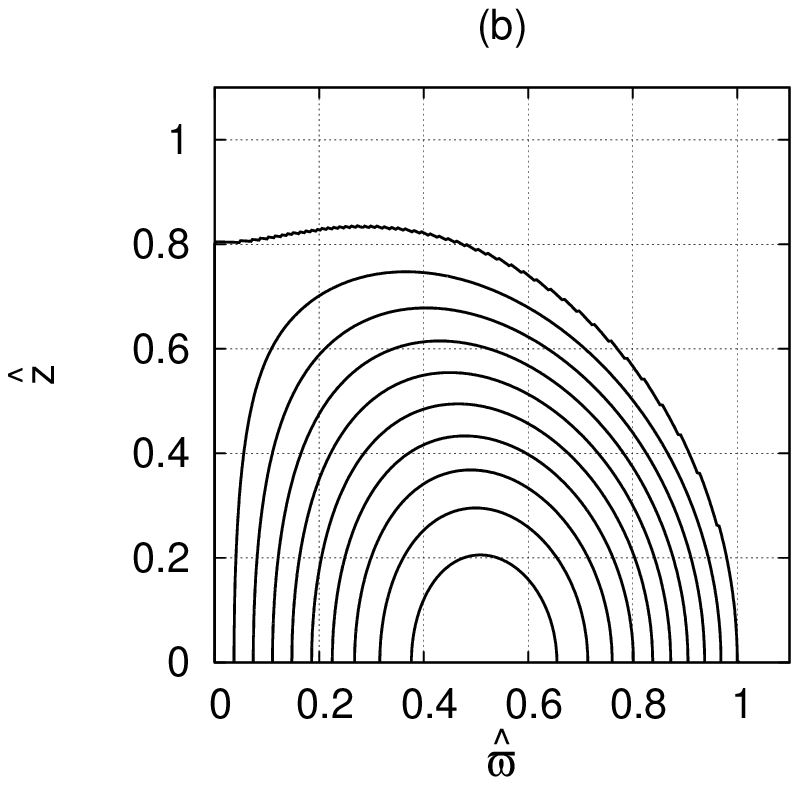}
 \\
\includegraphics[width=8.5cm]{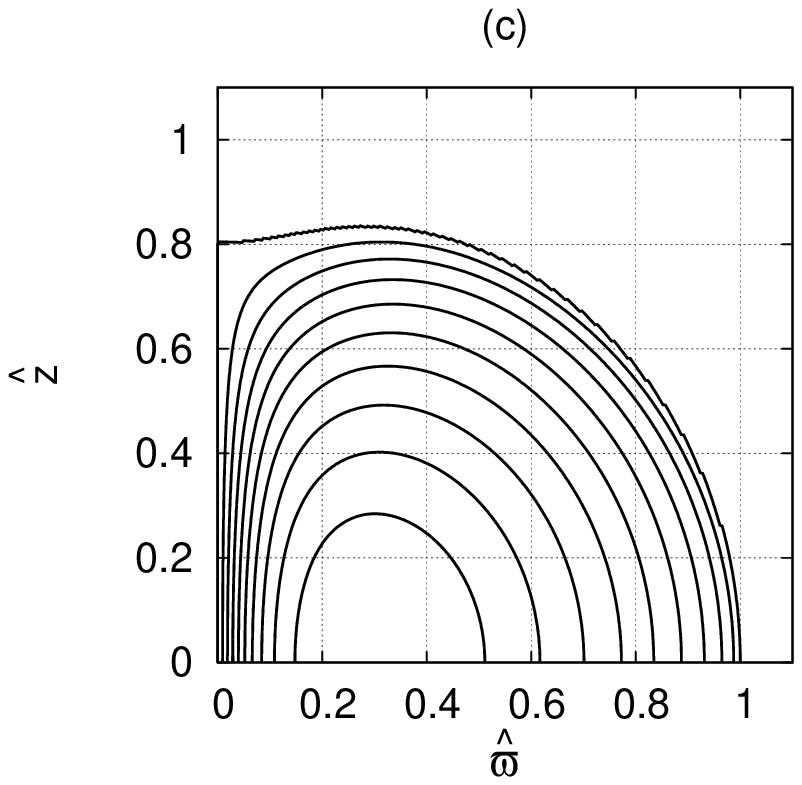} & 
\includegraphics[width=8.5cm]{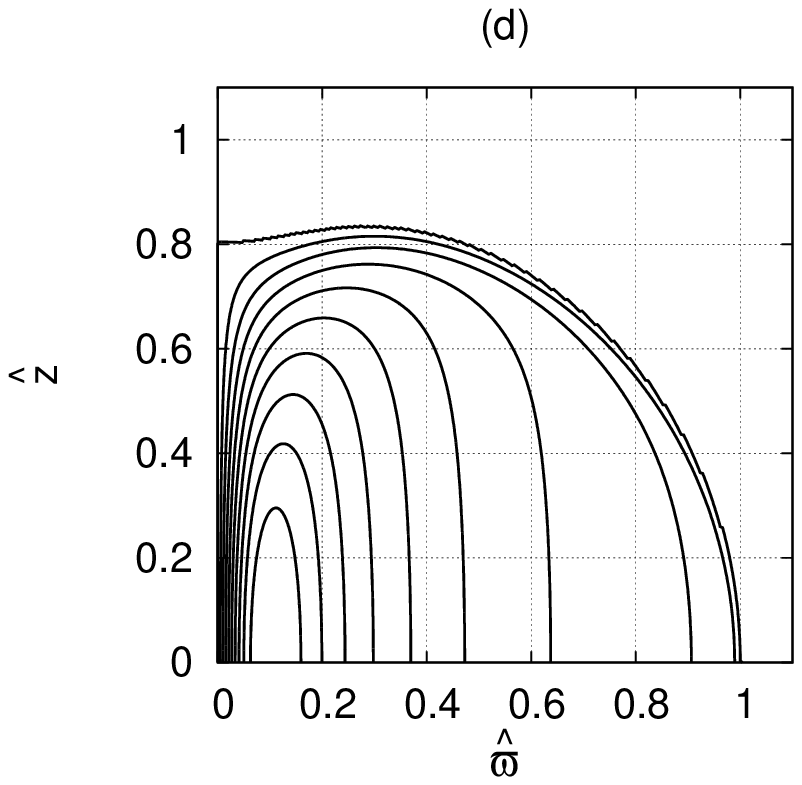} 
\end{tabular}

\caption{Contours of $\hat{\rho}$ and the toroidal magnetic field ($\hat{B}_\varphi$).
The difference between the two adjacent contours is 1/10 of the 
maximum value of $\hat{\rho}$ or $\hat{B}_\varphi$.
(a) The contour of a  $\rho$ with non-magnetic field.
(b) The contour of a $\hat{B}_\varphi$ of {rigid} toroidal field with $k=1$.
(c) The contour of a $\hat{B}_\varphi$ of type 1 {differential} toroidal field with $\hat{h}=0.1$.
(d) The contour of a $\hat{B}_\varphi$ of type 2 {differential} toroidal field with $\hat{h}=0.1$. 
}
\label{Fig:contours}
\end{figure*}

First, we will compare one non-magnetized equilibrium state and
six magnetized equilibrium states with {\it rigid} or {\it differential}
toroidal fields.  The numerical models and results are 
tabulated in table \ref{Tab:tab1} and plotted in figures \ref{Fig:profiles} and
\ref{Fig:contours}. 
The relative value of the Virial relation (VC) in all solutions is very small (see accuracy verification in Appendix A).
We calculated many magnetized equilibrium states and displayed some solutions
whose ratios of the rotational energy to magnetic energy are equal (${\cal M} / T = 1.0 \times 10^{-2}$). 
This energy ratio is smaller than Fu \& Lai's criterion (${\cal M}/T \sim 2.0 \times 10^{-1}$),
but larger than that of \cite{Muhlberger_et_al_2014} (${\cal M}/ T \sim 5.6 \times 10^{-3}$).
These toroidal magnetic fields could suppress the low-$T/|W|$ instability. 

Since the values of $\beta$ for these solutions are almost the same, their
magnetic energies are also almost the same. As seen in Table \ref{Tab:tab1},  
however, the ratio $B_{\max} / B_{\rm ave}$ is different for each model.  
The ratio for the {\it rigid} toroidal solution with $k=1$ is 1.88. 
This value is almost as the same as those for the type 1 
{\it differential} toroidal solution with $\hat{h} = 0.1$ ($1.70$) or the type 2
{\it differential} toroidal solutions with $\hat{h} = 0.1$ (1.8) and 
$\hat{h} = 0.2$ (1.99). On the other hand, 
the ratios of the {\it rigid} toroidal solution with $k=2$ 
and the type 2 {\it differential} toroidal solution with $\hat{h} = 0.3$ 
are 2.48 and 2.77, respectively. These larger values indicate that 
the toroidal fields of these two models 
are more localized than those of other solutions.

The profiles of the toroidal magnetic fields are different and rely heavily 
on their type.
Fig. \ref{Fig:profiles} displays the profiles of $d \hat{\varpi} \hat{\Omega} / d \hat{\varpi}$ (left-hand panel)
and the re-dimensionalized $B_\varphi$ (right-hand panel). The curves represent the following:
(i) the thick solid curve shows a type 2  {\it differential} field with $\hat{h} = 0.1$, 
(ii) the thick dashed curve shows a type 2 {\it differential} field with $\hat{h} = 0.2$, 
(iii) the thick dotted curve shows a type 1 {\it differential} field with $\hat{h} = 0.1$, 
(iv) the thin solid curve shows a {\it rigid} field with $k = 1$ and
(v) the thin dashed curve shows a {\it rigid} field with $k = 2$.
The magnetic winding by the differential rotation 
($d \hat{\varpi} \hat{\Omega} / \hat{\varpi}$ in the left-hand panel) 
reaches a maximum value at the rotational axis $\hat{\varpi} = 0$
and the value decreases rapidly near the centre. On the other hand,
the profiles of the {\it rigid} toroidal fields reach maximum values in the middle regions 
[$\hat{\varpi} \sim 0.5$ with $k=0$ (curve iv) and  $\hat{\varpi} \sim 0.6$ with $k=1$ (curve v)].
These distributions do not represent the magnetic winding by the differential rotation
($d\hat{\varpi}\hat{\Omega} / d \hat{\varpi}$ in the left-hand panel).
The profile of the type 1 {\it differential} toroidal field is gradual. 
The ratio of $B_{\max} / B_{\rm{av.}}$ is the smallest among these seven solutions.
This functional form of $I$ corresponds to $B_\varphi$-constant 
toroidal fields such as the $v$-constant rotation law, because
the functional form of the type 1 {\it toroidal } field is $I \sim \varpi (B_\varphi = $ const.$)$  
when the star is incompressible.

In contrast, the value of the type 2 {\it differential} toroidal fields reach 
a maximum near the rotational axis [$\hat{\varpi} \sim 0.2$ with $\hat{h} = 0.2$ (curve ii)
and $\hat{\varpi} \sim 0.1$ with $\hat{h} = 0.1$ (curve i)]. 
As the value of $\hat{h}$ decreases, the degree of the {\it differential} 
toroidal field becomes large and the maximum point reaches the rotational 
axis. These type 2 {\it differential} toroidal fields resemble
the magnetic winding by the differential rotation.
In particular, the toroidal field with $\hat{h} = 0.1$ closely matches 
the magnetic winding (fig. \ref{Fig:profiles}). Therefore, we can obtain 
the toroidal field wound up by differential rotation 
by using type 2 {\it differential} toroidal fields
under the assumption that initial profile of the magnetic 
field is uniform (equation \ref{Eq:Bt_wound}) .

Fig.  \ref{Fig:contours} shows the contour maps 
of $\hat{\rho}$ and $\hat{B}_\varphi$ for these solutions.
Panel (a) displays the density 
distribution of the non-magnetized solution.
Since the magnetic energy of our solutions is
smaller than the gravitational energy 
(${\cal M}/ |W| = {\cal M} / T \times T / |W| \sim 10^{-4} $),
all solutions in table \ref{Tab:tab1} have similar isopycnic surfaces.   
The other panels show distributions of the toroidal fields.
Panel (b) shows the contour of a {\it rigid} toroidal field with $k = 1$.
This configuration is the same as those in previous works
(e.g., see fig.  2 in \citealt{Lander_Jones_2009}).
Panels (c) and (d) display a type 1 {\it differential}
field with $\hat{h} = 0.1$ and a type 2 {\it differential}
field with $\hat{h} = 0.1$, respectively. As we see in fig.  \ref{Fig:profiles},
the toroidal magnetic fields of a type 2 {\it differential} field 
are concentrated and localized near the rotational axis.
 
\subsection{Highly localized {\it differential} toroidal magnetic field}

\begin{table*}
\begin{center}
\caption{Numerical results for each model.}
 \begin{tabular}{cccccccccc}
\hline  
 $q$ & $\hat{I}_0^2$ &  $\hat{h}$  &$\beta$  & ${\cal M} / T$ & $B_{\max}$ (G)
& $B_{\max} / B_{\rm ave}$ &  $V_{A\max}$ $(\mathrm{cm/s})$ & VC \\
\hline
&&&&& \hspace{-100pt} {\it Rigid} toroidal \\
\hline
0.8 & 7.12E-3 & - & 4.02E-2 & 1.00E-2 & 3.67E16 & 1.88 & 4.14E8 & 1.12E-5 \\
\hline
&&&&& \hspace{-100pt} Type 2 {\it differential } toroidal \\
\hline
0.8 & 1.00E-3 & 0.5 & 4.02E-2 & 9.44E-3 & 3.31E16 & 1.75 & 3.32E8 & 1.12E-5\\
0.8 & 5.00E-4 & 0.4 & 4.02E-2 & 8.25E-3 & 3.09E16 & 1.75 & 2.94E8 & 1.12E-5\\
0.8 & 2.00E-4 & 0.3 & 4.01E-2 & 6.05E-3 & 2.71E16 & 1.79 & 2.48E8 & 1.12E-5\\
0.8 & 5.00E-5 & 0.2 & 4.01E-2 & 2.91E-3 & 2.08E16 & 1.98 & 1.86E8 & 1.12E-5\\
0.8 & 3.50E-6 & 0.1 & 4.01E-2 & 4.21E-4 & 1.10E16 & 2.76 & 9.84E7 & 1.12E-5\\
0.8 & 1.50E-7 & 0.05& 4.00E-2 & 2.83E-5 & 4.56E15 & 4.41 & 4.07E7 & 1.11E-5\\
\hline
 \end{tabular}
\label{Tab:tab2}
\end{center}
\end{table*}

\begin{figure*}
 \includegraphics[width=8.5cm]{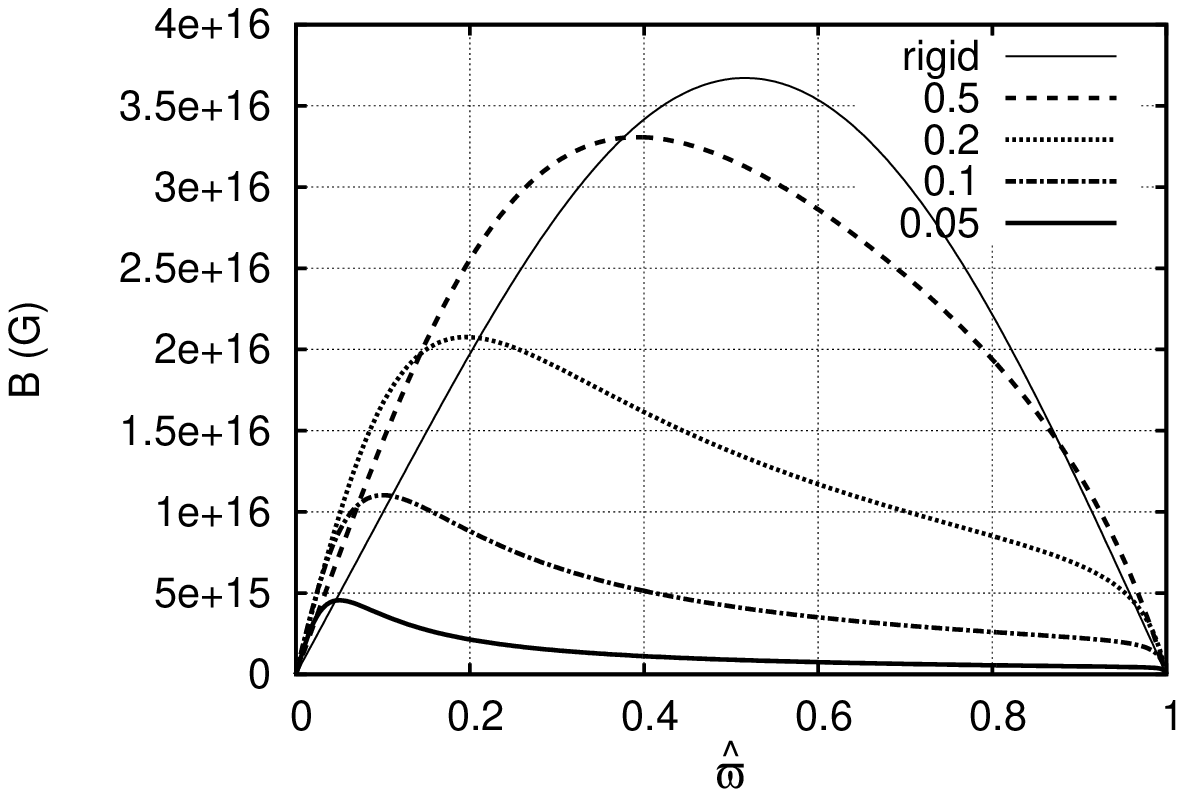}
 \includegraphics[width=8.5cm]{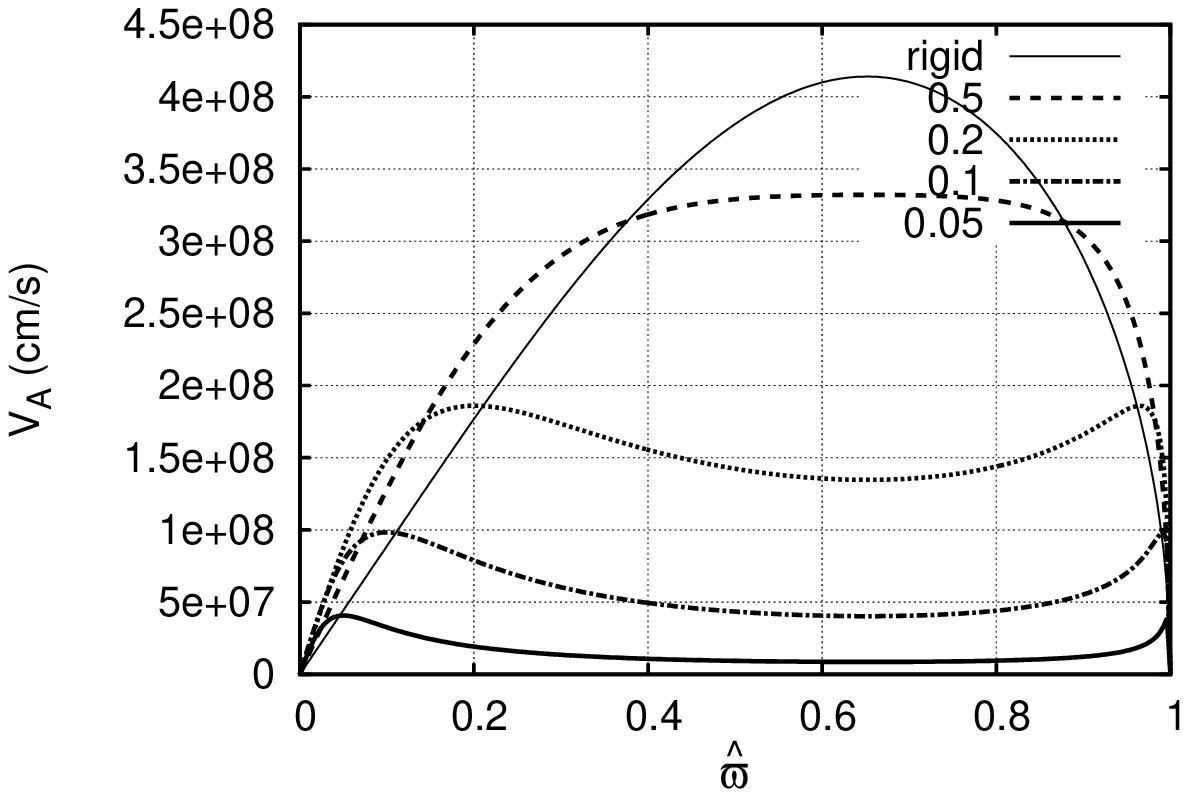}

\caption{Left: the profiles of $B_\varphi$ on the equatorial plane.
Each curve denotes a {rigid} field with $k=1$ (thin solid), 
and a type 2 {differential} field with $\hat{h} = 0.5$ (thick dashed), 
$\hat{h} = 0.2$ (thick dotted), $\hat{h} = 0.1$ (thick dot-dashed) and $\hat{h} = 0.05$ (thick solid).
Right: the profiles of $V_A$ on the equatorial plane. 
Each curve denotes a  {\it rigid} field with $k=1$ (thin solid), 
and a type 2 {\it differential} field with $\hat{h} = 0.5$ (thick dashed), 
$\hat{h} = 0.2$ (thick dotted), $\hat{h} = 0.1$ (thick dot-dashed) and $\hat{h} = 0.05$ (thick solid).}
\label{Fig:Bt_r}
\end{figure*}

Next, we calculated the type 2 {\it differential} toroidal field by changing the 
value of $\hat{h}$ and compared the resulting models and the {\it rigid} toroidal field model.
We show one {\it rigid} toroidal model and six {\it differential} toroidal models.
Since $\hat{h}$ determines the degree of the {\it differential} toroidal 
magnetic field, the toroidal magnetic field is localized and concentrated
near the centre when the value decreases. We calculated many equilibrium states by
changing the value of $\hat{h}$ and $\hat{I}_0$, and obtained one solution sequence.
The numerical results of the solution sequence are tabulated in Table \ref{Tab:tab2} and
the profiles of the re-dimensionalized toroidal  magnetic fields 
and the Alfv\'en velocity are plotted in fig. \ref{Fig:Bt_r}. 
In this sequence, the local maximum values 
of the {\it differential } toroidal magnetic fields are comparable to those
of a {\it rigid} field at that point (see fig. \ref{Fig:Bt_r}).

As seen in fig. \ref{Fig:Bt_r}, 
the location of the maximum point approaches the centre 
as the value of $\hat{h}$ decreases. This is  because the degree of the 
{\it differential} toroidal magnetic field increases. 
The maximum point of the {\it rigid} toroidal field is located at 
$\hat{\varpi} \sim 0.5$ and that point  in the {\it differential} toroidal fields becomes
$\hat{\varpi} \sim 0.4$ ($\hat{h} = 0.5$), $\hat{\varpi} \sim 0.2$ ($\hat{h} = 0.2$),
$\hat{\varpi} \sim 0.1$ ($\hat{h} = 0.1$) and $\hat{\varpi} \sim 0.05$ ($\hat{h} = 0.05$) as seen in 
the left-hand panel of fig. \ref{Fig:Bt_r}.
The profiles of Alfv\'en velocities have the same tendency in the right-hand panel.
The values of the ratio $B_{\rm max} / B_{\rm ave}$ increases as the degree of the 
{\it differential} increases.  The ratios are $1.75 (\hat{h} = 0.5)$, $1,98 (\hat{h} = 0.2)$, 
$2.76 (\hat{h} = 0.1)$ and (4.41 $\hat{h} = 0.05$). Thus, the {\it differential} toroidal 
magnetic field is localized and concentrated near the centre when the degree of the
{\it differential} is large. 

However, the magnetic field energy ratios (${\cal M}/ T$) of the {\it differential} toroidal field 
solutions are much smaller than that of the {\it rigid} one (Table \ref{Tab:tab2}). 
As fig. \ref{Fig:Bt_r} shows, the strength of the 
magnetic field and the Alfv\'en velocity
in each differential model near the local maximum point are almost  
the same as those of the {\it rigid} model at the same radius.
However, the magnetic energies of the differential models 
 are smaller than that of the {\it rigid} model
because of their localization and concentration near the point.
Although the magnetic energy ratio (${\cal M}/T$) of the 
{\it rigid} solution is $\sim 1.0 \times 10^{-2}$, 
the energy ratios of type 2 {\it differential} solutions with small $\hat{h}$ are 
$\sim 2.91\times 10^{-3}$ ($\hat{h} = 0.2$), $\sim 4.21 \times 10^{-4}$ ($\hat{h} = 0.1$),
 and $\sim 2.85 \times 10^{-5}$ ($\hat{h} = 0.05$).  
The magnetic energy of the type 2 {\it differential} toroidal solution 
with $\hat{h} = 0.05$ is much smaller than that of the {\it rigid}
toroidal model,  even if the strength of the toroidal magnetic field 
is comparable to that of the {\it rigid field } at the local maximum point $(\hat{\varpi}\sim 0.05)$.
The  localisation and concentration of the toroidal magnetic field 
near the rotational axis can be realized by using type 2 {\it differential} toroidal fields. 

\section{Discussion and concluding remarks}
  
\subsection{Suppression of the low-$T/|W|$ instability by low ${\cal M} / T$ magnetic fields}

The low-$T/|W|$ instability is related to 
the corotation point inside the star (\citealt{Andersson_2003}).
Since the wave pattern speed of an unsTable mode is equal to the rotational velocity of the background star, 
the corotation resonance would be driven at the point. 
Recent time evolutions of linear analysis have shown that the 
f mode becomes unsTable when the corotation point appears inside the star
(\citealt{Passamonti_Andersson_2015}).
The low-$T/|W|$ instability depends on the 
degree of the differential rotation (\citealt{Shibata_Karino_Eriguchi_2003}),
but the location of the corotation point is not obvious.
Linear analysis has also illustrated that the
location of the corotation point depends on both the  
degree of the differential rotation and the value of $\beta$ (\citealt{Passamonti_Andersson_2015}).
As the value of $\beta$ decreases and the degree of the differential 
rotation increases, the location approaches the rotational axis.
If the value of $\beta$ is 0.04 and the degree of the differential rotation
is $\hat{d} = 0.3$ as we have calculated in this paper, 
the corotation point will appear at $\hat{\varpi} \sim 0.27$ 
(see fig. 6 in \citealt{Passamonti_Andersson_2015}).
Therefore, the location of the corotation point will appear 
near the rotation axis in a differential rotating star.

The suppression of the toroidal magnetic field may 
depend on the local strength of the toroidal magnetic field
and Alfv\'en velocity, 
since the slow-magnetosonic points are determined locally 
by the values of density and toroidal magnetic fields at the point (\citealt{Fu_Lai_2011}). 
As seen in figures \ref{Fig:profiles} and \ref{Fig:Bt_r}, 
the maximum point of the {\it rigid} toroidal magnetic field is located
at $\hat{\varpi} \sim 0.5$ -- 0.6. These values are larger than 
the location of the corotation point ($\hat{\varpi} \sim$ 0.27).
On the other hand, the maximum point of the type 2 {\it differential} toroidal 
magnetic field is located at the inner region of the star.
The maximum point is located at $\hat{\varpi} \sim 0.2$ 
when the degree of the {\it differential} is $\hat{h} = 0.2$.
The location approaches the rotation axis as the 
degree of the {\it differential} toroidal magnetic field 
increases (fig. \ref{Fig:Bt_r}).
Since the {\it differential} toroidal magnetic field 
represents the magnetic winding by the differential rotation very well (fig. \ref{Fig:profiles}),
the profile of the toroidal magnetic field would become
that of the type 2 {\it differential} toroidal magnetic fields
if the toroidal field is wound up from the initial poloidal magnetic field
by the differential rotation.
If the star has a large differential rotation, 
the degree of the {\it differential} toroidal magnetic field
would also become large because the toroidal field is developed by the magnetic winding.
As a result, both the maximum point of the toroidal magnetic field
and the corotation point approach the rotational axis. 
We assume that both points appear near the
centre when the star has a large differential rotation. 
Such a {\it differential } toroidal field would suppress the low $T/|W|$ 
instability easily, even if the total magnetic 
field energy ratio (${\cal M} / T$) were much smaller than that of the {\it rigid} toroidal field,
because the strength of the local toroidal magnetic field 
and Alfv\'en velocity are comparable to those
of the {\it rigid} toroidal field (fig. \ref{Fig:Bt_r}).

Such low-${\cal M}/T$ suppression might be realized by the type 2 
{\it differential } toroidal field. Since \cite{Muhlberger_et_al_2014}
calculated both rotation and magnetic winding simultaneously, 
the distribution of their toroidal magnetic 
field would be type 2 {\it differential} rather than
{\it rigid}. The suppression of the low-$T/|W|$ instability might occur 
in a {\it differential} magnetic field configuration even if the magnetic energy ratio
is very small. On the other hand, \cite{Fu_Lai_2011} adopted a
simple {\it rigid} toroidal magnetic field solution as a 
background model for their linear analysis. 
The suppression of the low-$T/|W|$ instability might need
larger toroidal magnetic field energy than is available from  a {\it differential} field
because the maximum point of the toroidal field 
is located at the outer region of the 
corotation point. 

The suppression would depend on the degree of the 
{\it differential} toroidal field as the low-$T/|W|$ instability depends
on the degree of the differential rotation. 
In order to study the low-${\cal M}/T$ suppression systematically, in the future, 
we will perform linear analysis using the {\it differential} toroidal
field obtained here. Such work may  explain the difference in the energy ratio (${\cal M}/T$)
between a linear analysis with a {\it rigid} toroidal field  (\citealt{Fu_Lai_2011}) 
and a  numerical simulation with a {\it differential} toroidal field 
made by magnetic winding (\citealt{Muhlberger_et_al_2014}).  

\subsection{Concluding remarks}

We investigated new magnetized equilibria with 
differential rotation and {\it differential} toroidal fields in this paper.
We developed new functional forms for {\it differential} toroidal fields in
analogy to the rotation law of the differential rotation
and obtained equilibrium states self-consistently 
by using a self-consistent numerical scheme.

The {\it differential} toroidal magnetic fields have two functional forms.
The functional form of the type 1 {\it differential} toroidal field 
is similar to the $v$-constant differential rotation law.  
The profile of the toroidal field is a gradual slope (fig. \ref{Fig:profiles}).
Since the toroidal magnetic field becomes constant 
when the star is incompressible, the type 1 
{\it differential} toroidal magnetic field 
is analogous to a $B_\varphi$-constant toroidal field law. 
The type 2 {\it differential} toroidal field  
is similar to the $j$-constant rotational law. 
Such {\it differential} toroidal fields represent the 
magnetic winding by the $j$-constant differential rotation very well. 
The profiles of the toroidal magnetic fields represent the magnetic winding by the 
differential rotation (fig. \ref{Fig:profiles}).
The type 2 {\it differential } toroidal field is localized and 
concentrated near the rotational axis when the degree of the
{\it differential} toroidal magnetic field is large (fig. \ref{Fig:contours}). 

The low-$T/|W|$ instability would be suppressed by the 
concentrated and localized {\it differential} toroidal fields 
more efficiently than a {\it rigid} toroidal field.
The maximum strength of the 
{\it differential} toroidal magnetic field can be locally comparable 
to that of a {\it rigid} toroidal magnetic field, 
even if the total magnetic energy of the {\it differential}
field is much smaller than that of a {\it rigid} one (fig. \ref{Fig:Bt_r}).
 Since the low-$T/|W|$ instability and the suppression by the toroidal magnetic 
field would depend on the strength of the local rotational velocity 
and the toroidal magnetic field, 
the localized and concentrated {\it toroidal} magnetic field 
is efficient for the suppression even if the total 
magnetic energy is small. 
Such low-${\cal M}/T$ suppression would be realized by the 
{\it differential} toroidal field wound up from the initial 
poloidal magnetic field by differential rotation. 
The {\it differential} toroidal magnetic field 
plays a key role in explaining the difference in the energy ratio 
(${\cal M}/ T$) between a linear analysis with a {\it rigid} toroidal field (\citealt{Fu_Lai_2011})
 and a numerical simulation with a {\it differential} toroidal field wound 
up from the initial poloidal magnetic field 
by differential rotation (\citealt{Muhlberger_et_al_2014}).

\section*{Acknowledgements}

The author would like to thank the anonymous reviewer for useful comments.
K.F. would like to thank the members of the GG seminar and H. Okawa for
valuable and useful comments.
K.F. is supported by Grant-in-Aid 
for Scientific Research on Innovative Areas, 
no. 24103006.

\bibliographystyle{mn}


 \appendix

\section{Accuracy verification}

\begin{figure}
 \begin{center}
\includegraphics[width=9cm]{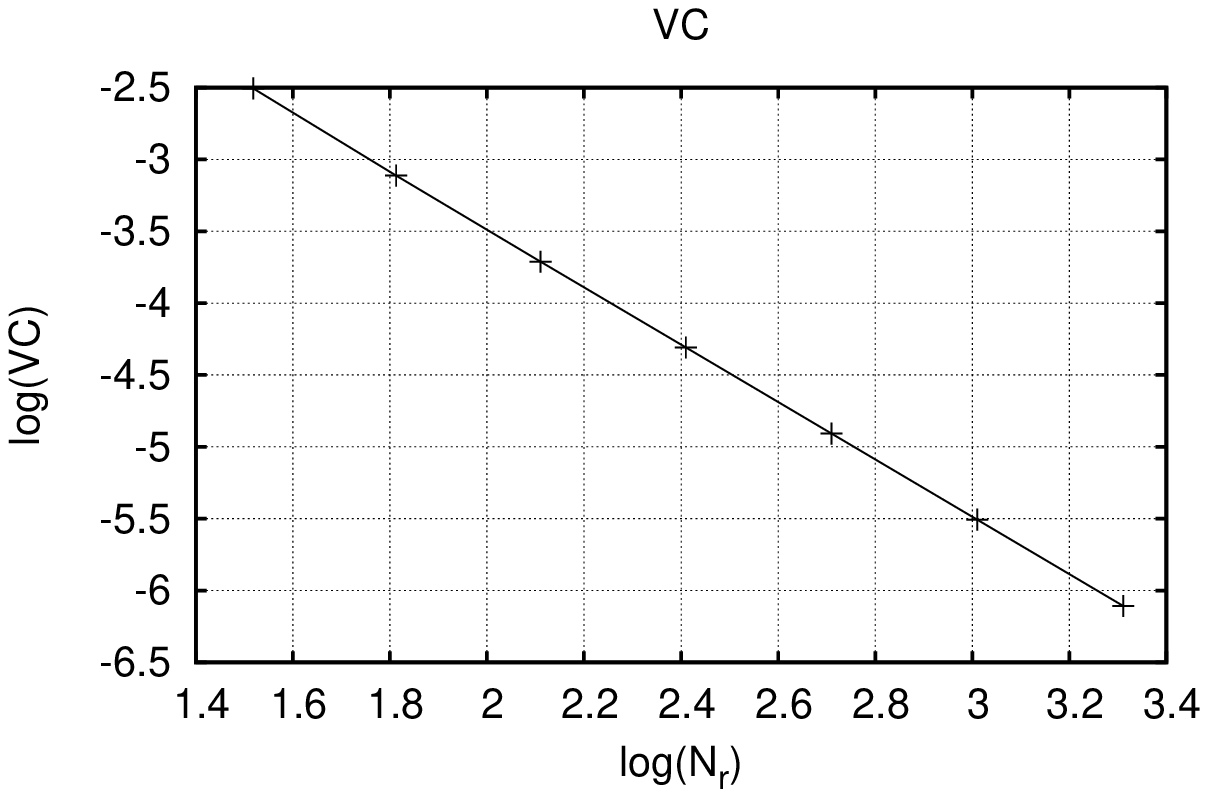}
 \end{center}
\caption{The Virial quantity VC, plotted against the number of grid points
in the $r$-direction.}
\label{Fig:VC}
\end{figure}

To check the numerical accuracy, 
we computed a relative value of the Virial relation as follows:
\begin{eqnarray}
 { \rm VC} \equiv \frac{|2T + W + 3 \Pi + {\cal M}|}{|W|}.
\end{eqnarray}
Our numerical domain is defined as $0 \leq \hat{r} \leq 1$ in the radial 
direction and $0 \leq \theta \leq \pi / 2$ in the angular direction.
Fig. \ref{Fig:VC} displays the value of VC as a function of the 
number of grid points in the $r$-direction.
To plot this graph, we computed 
the solutions with $j$-constant differential rotation and 
a type 2 {\it differential} toroidal field ($\hat{h}$ = 0.3) 
changing the number of grid points in the $r$-coordinate, but fixing the number of 
grid points in the $\theta$-direction as $n_{\theta} = 513$ (\citealt{Fujisawa_Yoshida_Eriguchi_2012}).
The value of VC decreases in proportion to the square inverse of the number of grid points
(\citealt{Lander_Jones_2009}; \citealt{Fujisawa_Yoshida_Eriguchi_2012}),
because we adopted Simpson's scheme as an  integral method in our numerical code. 
The typical value of VC is smaller than $10^{-4}-10^{-5}$
when the mesh numbers were $N_r = 513$ and $N_\theta = 513$.
This value is sufficiently small and the system 
converges very well (cf. \citealt{Hachisu_1986a}; \citealt{Tomimura_Eriguchi_2005}; 
\citealt{Lander_Jones_2009}; \citealt{Fujisawa_Eriguchi_2013}).
Thus, we fixed $N_r = 513$ and $N_\theta = 513$ in all calculations 
in this paper.

\end{document}